\documentclass[preprint,proceedings]{rmaa}

\usepackage{paralist}

\usepackage{psfrag,color}

\def\Msun{\hbox{$\rm\thinspace M_{\odot}$}}

\SetYear{2002}
\SetConfTitle{Galaxy Evolution: Theory and Observations}

\title{Galaxy Formation: Clues from the Milky Way}

\author{
  G. Gilmore\altaffilmark{1}} 

\altaffiltext{1}{Institute of Astronomy, Cambridge, U.K.}

\shortauthor{Gerry Gilmore}
\shorttitle{Milky Way Formation}

\fulladdresses{
\item G. Gilmore, Institute of Astronomy, Madingley Road, Cambridge CB3
0HA, U.K.}

\listofauthors{G. Gilmore}
\indexauthor{Gilmore, G.}

\abstract{Stellar Populations are the fossil record of Galactic
evolution. Interpretation of this record in the Local Group allows one
to determine 
reliably the dominant physics controlling the evolution of those
galaxies which are typical of the luminosity in the Universe, and is an
essential prerequisite to understanding necessarily limited 
data at high redshifts. In our Galaxy, the key issues are the
places and times of formation and merger of the oldest stellar
populations : the halo, thick disk and bulge - and their overlaps and
evolutionary relationships, if any. New results on studies of the
stellar initial mass function at high redshift, the stellar
populations of the Galactic bulge, and the merger history of the
Galactic disk are reviewed.
}

\addkeyword{galaxies: Milky Way}
\addkeyword{stars: Initial Mass function}
\addkeyword{Galaxy formation}

\begin{document}
\maketitle

\section{Introduction}

\begin{figure}
\label{fig1}
\includegraphics[width=\columnwidth]{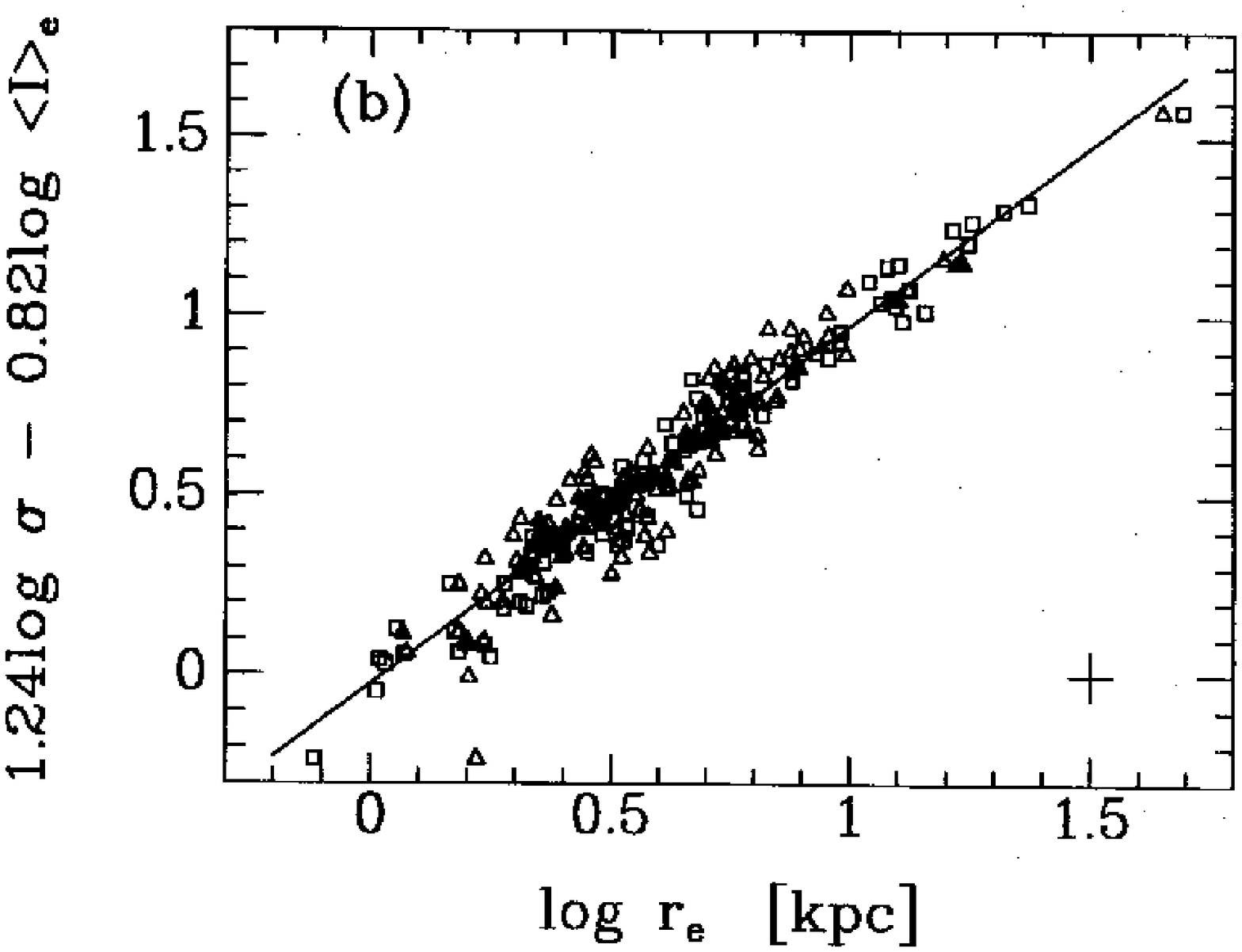}
\vskip10truecm
\caption{Top: Edge-on
view of the fundamental plane of spheroids, from J\o rgensen et al.\ (1996). 
{\em Boxes:} Ellipticals. {\em Triangles:} Bulges (S0s). Typical
error bars are shown. LOWER
 A plot of [Fe/H] (filled squares) or [O/H] - 0.37
 vs absolute V-band magnitude. The
dotted line is a rough fit to the [Fe/H]-MV relation for the dSph and
transition objects. Sagittarius corresponds to the points near
(MV,[Fe/H]) ~ (-13.4,-1.0). Square symbols refer to dSph or dE
galaxies; triangles refer to transition galaxies (denoted dIrr/dSph in
Table 1); circles refer to dIrr systems. Filled symbols correspond to
[Fe/H] abundances determined from stars, while open symbols denote
oxygen abundance estimates from analyses of HII regions and planetary
nebulae. [From Mateo 1998.] These systematic relations, with little
scatter for galaxies without active current star formation,
show that the galaxies
of the Local Group should be fair samples of galaxies in general.}
\end{figure}

We are fortunate that the Milky Way itself, and more generally the
galaxies of the Local Group, appear typical of the `mean' population
in the Universe. Detailed studies of the systematic properties of
galaxies, leading eventually to the concept of the `fundamental
plane', have shown that galaxies form well-defined families (Figure
\ref{fig1}). Thus, we may immediately deduce that galaxy formation models
should produce `Local Group-like' and `Milky Way-like' systems without
special effects. Conversely, detailed study of the Local Group can
contribute to the general study of galaxy formation and evolution, and
the nature and distribution of dark matter on small scales.

\begin{figure*}[t!]
\label{fig2}
\vskip 20truecm
\caption{Three views of the young, star-forming Galactic Bulge: TOP:
the 7micron ISOGAL 
image. Bright features in the image indicate star formation and/or
young stellar regions; MIDDLE: a SCUBA map from Pierce-Price etal 2001,
highlighting regions of active continuing star formation, and
illustrating the considerable molecular gas reservoir; LOWER: the
CHANDRA view, dominated by hot gas and supernova activity.} 
\end{figure*}

Topics of relevance here where local studies are proving especially
significant include the nature of the Galactic Bulge: what is its age,
abundance and assembly history? Stellar studies locally of course are
of critical general significance: only locally can we determine the
stellar Initial Mass Function directly in a wide range of
environs. This function directly controls the chemical and luminosity
evolution of the Universe. Galactic satellite galaxies are proving
the most suitable environs to quantify the nature and distribution of
dark matter, and  to test the small scale predictions of hierarchical
galaxy formation models. The Galactic disk itself, of course, is a key
test of angular momentum distributions, chemical evolution, and
merger histories.

\section{ The Galactic Bulge}

The Milky Way galaxy provides a unique opportunity to learn about the
formation, the structure and the evolution of galaxies. The central
parts of the galactic bulge and disk have remained elusive, though,
due to the extremely high extinction at short wavelengths and poor
spatial resolution at longer wavelengths (Figure \ref{fig2}). 
Most of the current belief
that the stellar content of the galactic bulge is old, $\ga 10$ Gyr,
and metal-rich, $[M/H]\sim$solar, results from studies in low
extinction regions (e.g.\ Baade's Window) at galacto-centric radii
$R>500$ pc (eg Rich 1998a). With the advent of infrared (IR) cameras and
adaptive optics techniques, the exploration of the galactic centre has
revealed the presence of massive stars that indicate recent star
formation (Genzel et al.\ 1994). Yet data concerning the relationship
between the central parsec of the galaxy and the Bulge, halo and disk
remains scarce.

There are clear similarities and distinctions between fundamental
properties of the different Galactic stellar populations, such as age,
metallicity, star formation history, angular momentum (figure~3)
which allow study of their individual histories. This is arguably one
of the greatest advantages of studies of Local Group galaxies: one is
able to disentangle the many different histories which have led to a
single galaxy.

\begin{figure}
\label{fig3}
\includegraphics[width=\columnwidth]{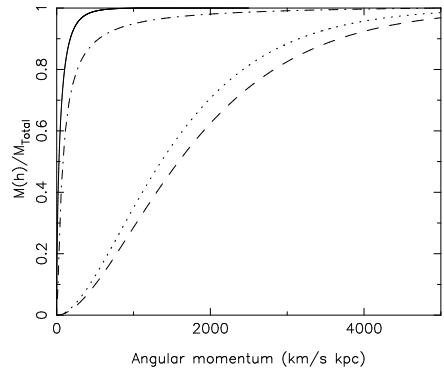}
\caption{Galactic populations angular momentum cumulative distribution
functions, showing bulge/halo and thin/thick disk
dichotomy. This indicates disparate evolutionary histories.}
\end{figure}

What we observe today is the time-integrated history of star formation, gas
flows, and mergers in the galaxy, and it may be envisaged that the formation
of the different components of the galaxy --- halo, Bulge, nucleus and disk
--- are not independent events. Common preconceptions about the Bulge being an
old, metal-rich small elliptical galaxy are being challenged (Wyse et al.\
1997). For instance, recent near-IR photometry and spectroscopy of stars in
the inner Bulge ($R\la500$ pc) suggest the presence of an intermediate-age
population ($t\sim1$ to 2 Gyr: e.g.\ Frogel 1999a). Did these stars form in
the Bulge, or in the nucleus? Is there any connection between the star
formation history and the formation of galactic structures such as a bar
(Blitz et al.\ 1993) or tri-axial Bulge (Nakada et al.\ 1991)?

The galactic bulge is fundamentally typical of all bulges in late-type spirals
(Frogel 1990). In particular, it is very similar to that of M31 and M32
(Davies et al.\ 1991; DePoy et al.\ 1993; Davidge 2000b, 2001; Rich 2001)
 and the central nuclei of M33 (Mighell \& Rich 1995;
Mighell \& Corder 2002; Stephens \& Frogel 2002) and NGC 247
and NGC 2403 (Davidge \& Courteau 2002), which all seem predominantly old and
metal-rich but most of which do contain bright AGB stars and possibly even
younger populations. Rich \& Mighell (1995) ask the question why the
integrated light of the bulge of M31 is so red despite the presence of an
intermediate population. This is probably due to the fact that the integrated
light results mainly from the red giants of $\sim$solar mass, and in addition
from the many red dwarfs that have been formed in any generation of stars.
This explains why observations of distant bulges show an old, metal-rich
content, whereas observations of spatially resolved stellar populations in
nearby bulges increasingly show the presence of younger as well as
metal-poorer stellar populations --- see also Lamers et al.\ (2002) for the
bulge of M51 and Rejkuba et al.\ (2001) for the giant elliptical NGC5128.

A major recent study of the Galactic bulge, vastly less affected by
forground dust extinction than all earlier studies, has recently been
completed by the ISOGAL consortium (van Loon etal 2002).

\begin{figure}
\label{fig4}
\includegraphics[width=\columnwidth]{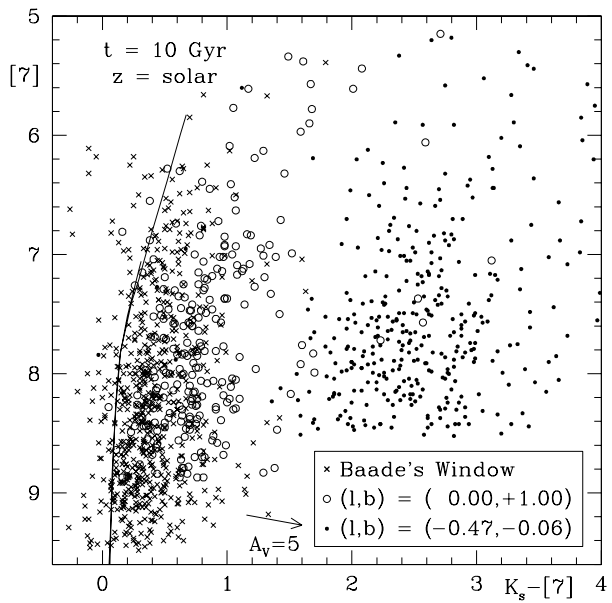}
\caption{ISOGAL mid-IR colour-magnitude data, illustrating the ability
to distinguish extinction from intrinsic stellar properties. Many of
the brightest sources here are intermediate-mass OH/IR and luminous
AGB stars. ISOGAL provides for the first time an opportunity to define
the spatial distribution of intermediate age stars, with masses
greater than $\sim1\Msun$.  From van Loon etal 2002 }
\end{figure}

Near- and mid-IR survey data from DENIS and ISOGAL were used to
investigate the structure and formation history of the inner
$10^\circ$ (1.4 kpc) of the Milky Way galaxy. Synthetic bolometric
corrections and extinction coefficients in the near- and mid-IR were
derived for stars of different spectral types, to allow the
transformation of theoretical isochrones into observable
colour-magnitude diagrams.
The observed IR colour-magnitude diagrams
could then be used to derive the extinction, metallicity and age for
individual stars (Figure~4). The inner galaxy is dominated, as expected, by an
old population ($\ga7$ Gyr). In addition, an intermediate-age
population ($\sim200$ Myr to 7 Gyr) was detected, which is consistent
with the presence of a few hundred Asymptotic Giant Branch stars with
substantial mass loss. Furthermore, young stars ($\la200$ Myr) are
found across the inner Bulge. The metallicities of these stellar
population components could also be derived (Figure~5).

These results can be interpreted in terms of an early epoch of intense
star formation and chemical enrichment which shaped the bulk of the
Bulge and nucleus, and a more continuous star formation history which
gradually shaped the disk, perhaps also involving accretion of
sub-solar metallicity gas from the halo. A possible increase in star
formation $\sim200$ Myr ago might have been triggered by a minor
merger. Ever since the formation of the first stars, mechanisms have
been at play that mix the populations from the nucleus, Bulge and
disk. Luminosity functions across the inner galactic plane indicate
the presence of an inclined (bar) structure at ${\ga}1$ kpc from the
galactic centre, near the inner Lindblad resonance. The innermost part
of the Bulge, within $\sim1$ kpc from the galactic centre, however
seems azimuthally symmetric. This is inconsistent with standard galaxy
bar models from the literature.

\begin{figure}
\label{fig5}
\includegraphics[width=\columnwidth]{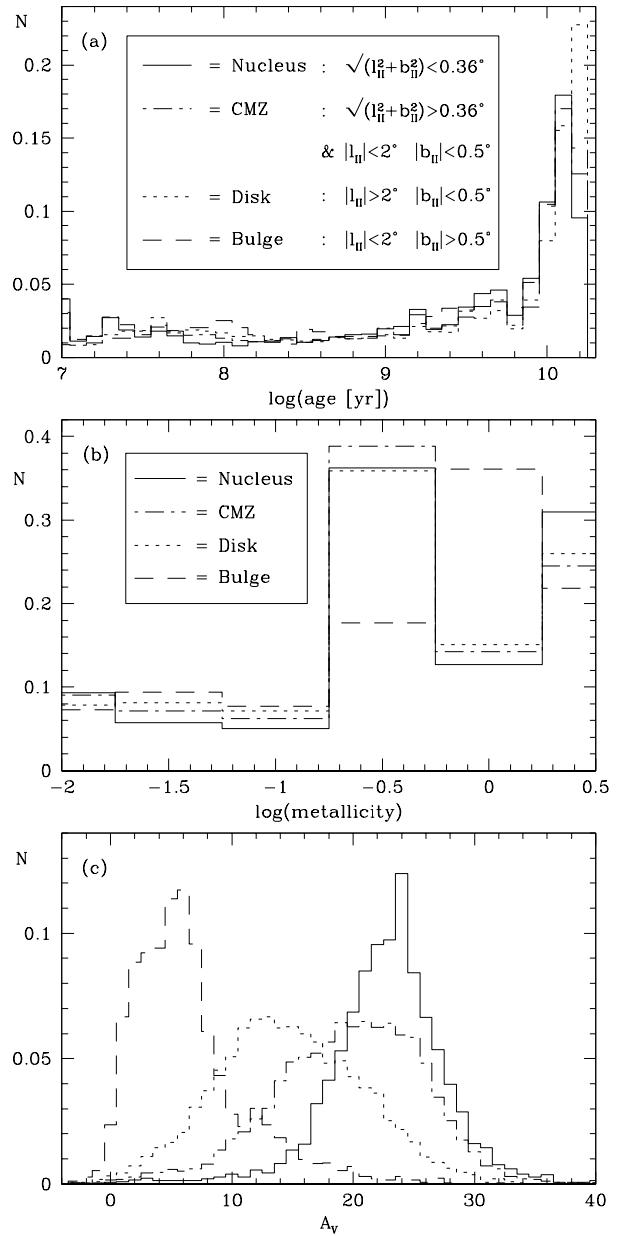}
\caption{Analysis of the ISO ISOGAL inner bulge mid-IR photometric
survey. The normalised (a) age (b) metallicity and (c) extinction
distributions  for the nucleus, central molecular zone, disk and
Bulge. From van Loon etal 2002}
\end{figure}

\section{The Stellar Initial Mass Function}

In addition to its possible relevance to dark matter problems, the
initial mass function (IMF) of low-mass stars in a wide variety of
astrophysical systems is of considerable intrinsic interest (see
e.g.~papers in Gilmore \& Howell 1998).  For example, the form of the
IMF at high redshift is of crucial importance for such aspects of
galaxy formation as the understanding of background light measurements
(e.g.~Madau \& Pozzetti 2000), galaxy luminosity and chemical
evolution.

Low-mass stars, those with main-sequence lifetimes that are of order
the age of the Universe, provide unique constraints on the Initial
Mass Function (IMF) when they formed.  Star counts in systems with
simple star-formation histories are particularly straightforward to
interpret, and those in `old' systems allow one to determine the
low-mass stellar IMF at large look-back times and thus at high
redshift. 

The faint luminosity function and low-mass IMF of stars formed
recently are in practise however difficult to establish; in very young
systems, one must often deal with the complexities of infrared
luminosity functions, while the interpretation of field star counts is
complicated by the errors in luminosity introduced by distance
uncertainties and unknown metallicity spreads.

In spite of that, recent heroic-scale efforts (cf.~Muench, Lada \&
Lada 2000), have quantified the local young IMF over three decades in
mass, extending well below the hydrogen-burning stellar mass limit.
Somewhat unexpectedly, the available data are consistent with a mass
function indistinguishable from that of the globular clusters (von
Hippel et al.~1996; Reid et al.~1999; Gilmore 2001; Kroupa 2002; but
see Eisenhauer 2001 for the view that the IMF does vary).  Thus, the
low-mass IMF is apparently invariant to first order with time and
metallicity. 

The high-mass IMF can be studied directly for current star formation,
and indirectly at high redshifts. The indirect method exploits the
sensitivity of the element-dependent yield of a Type~II supernova to
progenitor initial mass. Thus, observed element ratios in metal-poor
old stars may be interpreted to deduce the IMF slope for stars of the
enriching earlier generation, with mass $\ga 10 \Msun$. Analyses of
this type have been reported by several groups (eg Wyse and Gilmore
1988). A recent example of the type of data used is shown in figure~6.
The conclusion is that the high-mass metal-poor IMF 
was not very different, if at all, from that of stars forming today:
an approximately Salpeter IMF is deduced.

\begin{figure}
\label{fig6}
\includegraphics[width=\columnwidth]{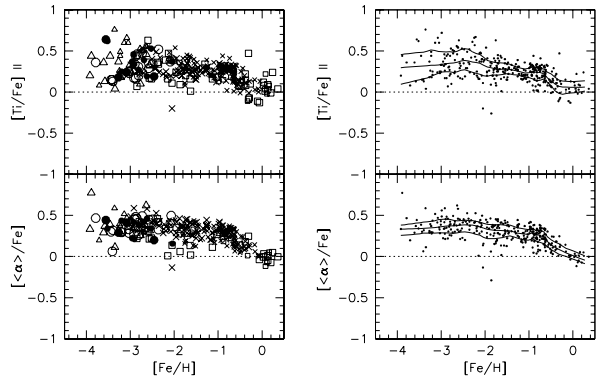}
\caption{Element ratios as a function of metallicity. The
approximately constant ratios for stars below [Fe/H]$\sim -1$ imply a
metal-poor hign mass IMF which is close to Salpeter in slope for all
abundances. Data from Carretta etal (2002). }
\end{figure}

\subsection{ The IMF in ancient environments}

A direct test of the invariance of the low mass stellar IMF to
environment at high redshifts is provided by comparison of the
faint stellar luminosity function in a dSph galaxy with that of a
stellar system that has similar stellar age and metallicity
distributions and which is known to contain no dark matter.  Empirical
comparison {of such} luminosity functions minimises the {need to use
the highly uncertain and metallicity-dependent} transformations
between mass and light (see D'Antona 1998 for a discussion of this
last point). Recently Wyse etal (2002) have reported the results of
such a direct comparison.

As a class, the dwarf spheroidal (dSph) companions of the Milky Way,
{defined by their extremely low central surface brightnesses and low
integrated luminosities (e.g.~Gallagher \& Wyse 1994)}, have internal
stellar velocity dispersions that are in excess of those expected if
these systems are in virial equilibrium, provided that their
gravitational potentials are provided by stars with a mass function
similar to that observed in the solar neighbourhood (see Mateo 1998
for a recent review).  The most plausible explanation {for} the
internal stellar kinematics {of these galaxies is} the presence of
gravitationally dominant dark matter, concentrated on small length
scales, leading to mass-to-light {ratios} a factor of ten to fifty
above those of normal old stellar populations.  The Draco dSph is
clearly dominated by an extended dark matter halo (Kleyna et
al.~2001).  This dark matter must be {cold to} be dominant on such
small scales ($\la 1$~kpc; cf.~Tremaine \& Gunn 1979; Gerhard \&
Spergel 1992; Kleyna et al.~2001).  Could some of the dark matter be
baryonic?  Low mass stars have high mass-to-light ratios; indeed stars
of mass 0.3~$M_\odot$ and metallicity one-hundreth of the solar value
-- of order the lowest mean metallicity measured for stars in dSph --
have V-band mass-to-light ratios of 24 in solar units (Baraffe et
al.~1997), and higher metallicity stars are even fainter.  Of course
faint stars could be viable dark matter candidates only if the stellar
initial mass function (IMF) in these systems were very different from
{the} apparently invariant IMF observed for other stellar systems,
such as the solar {neighbourhood} or globular clusters (cf.~Gilmore
2001).  

\begin{figure}
\label{fig7}
\includegraphics[angle=270,width=\columnwidth]{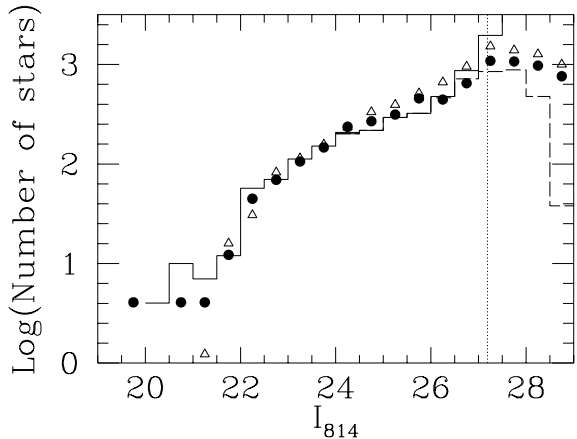}
\caption{The luminosity function of the UMi dSph galaxy, and that of
two old metal-poor globular clusters (M15, M92). The similarity shows
that the stellar IMF is not a function of environment. (from Wyse etal
2002.)}
\end{figure}

The stellar population of the Ursa Minor dwarf Spheroidal (UMi dSph)
is characterized by narrow distributions of age and of metallicity
(e.g.~Olszewski \& Aaronson 1985; Mighell \& Burke 1999; Hernandez,
Gilmore \& Valls-Gabaud 2000), with a dominant component that is
similar to that of a classical halo globular cluster such as M92 or
M15, i.e. old ($\ga 10$~Gyr) and metal-poor (mean [Fe/H] $ \sim
-2$~dex).  However, in contrast to globular clusters, {which have}
typical $(M/L)_V \la 3$ (e.g.~Meylan 2002), the internal dynamics
of the UMi dSph are apparently dominated by dark matter, since the
derived mass-to-light ratio is $(M/L)_V \ga 60$, based on the
relatively high value of its internal stellar velocity dispersion
(Hargreaves {et al.}  1994; see review of Mateo 1998).  

Faint star counts in the Ursa Minor dSph thus allow determination of
the low-mass IMF in a dark-matter-dominated external {galaxy in} which
the bulk of the stars formed at high redshift (a lookback time of 12
Gyr, the stellar age, corresponds to a redshift of $\ga 2.5$ for a
`concordance' Lambda-dominated cosmology; e.g.~Bahcall et al.~1999).

The results of Wyse etal are shown in figure~7. The main-sequence
stellar luminosity function of the Ursa Minor dSph, and {the} implied
IMF down to {$\sim 0.3$ M$_\odot$}, is indistinguishable from that of
the halo globular clusters M92 and M15, systems with the same old age
and low metallicity as the stars in the Ursa Minor dSph.  The
available {(indirect)} limits on the high-mass IMF, {inferred from}
elemental abundance ratios, are {also consistent} with the same IMF in
these two very different classes of systems. However, the globular
clusters show no evidence for dark matter, while the Ursa Minor dSph
is apparently very dark-matter dominated.

\section{ Merger histories, Merger participants}

Mergers and strong interactions between galaxies happen, and may well
be the dominant process in the determination of a galaxy's current
Hubble type, particularly in the context of modern
hierarchical-clustering theories of structure formation (e.g.~Silk \&
Wyse 1993).  The recently discovered (Ibata, Gilmore \& Irwin 1994)
Sagittarius dwarf spheroidal galaxy is inside the Milky Way
Galaxy, is losing a significant stellar mass through tidal effects
(Ibata et al.~1997), forming star streams in the halo (Mateo, Morrison
\& Olszewski 1998; Ibata et al.~2001; Yanny et al.~2000), but having
little effect on the present structure of the bulk of the Galactic disk.

In the standard hierarchical clustering and
merging picture of galaxy formation a thick disk is an expected
outcome of a significant merger. Depending on the mass, density
profile and orbit of the merging satellite, `shredded-satellite' stars
may retain a kinematic signature distinct from that part of the thick
disk that results from the heated thin disk.  Satellites on prograde
(rather than retrograde) orbits couple to the rotating
thin disk more efficiently, and thus a merger with such a system is
favored as the mechanism to form the thick disk (Quinn \& Goodman
1986; Velazquez \& White 1999).  If any kinematic trace of the
now-destroyed satellite galaxy is visible, it will be seen in the mean
orbital rotational velocity of stars.  The actual lag expected from
the shredded-satellite depends predominantly on the initial orbit and
the amount of angular momentum transport in the merger process, and is
not {\sl ab initio} predictable in a specific case.

The outcome of a merger of two stellar systems depends on several
factors, most importantly the mass ratio and density contrast. During
a merger, energy, momentum and angular momentum are re-distributed so
that the common aftermath of a merger between a large disk galaxy and
a smaller, but still significant, satellite galaxy (more massive than
the Sagittarius dwarf spheroidal galaxy) is a heated disk and a
disrupted satellite (Quinn \& Goodman 1986; Velaquez \& White
1999). This is currently the most plausible model for the origin of
the thick disk in our Galaxy (see reviews in Gilmore, Wyse \& Kuijken
1989; Majewski 1993) and those of other galaxies; the stochastic
nature of the merger process allows for a wide variety of, and indeed
non-existence of, thick disks in external galaxies, as observed,
provided only a small number of merger events are involved.

In general however, excluding special initial conditions, such as a
circular orbit at large distance (see Walker, Mihos \& Hernquist
1996), satellite debris stars will be on orbits characterised by lower
net rotational streaming about the Galactic Center than that of the typical 
scattered former thin-disk star at a given distance from the Galactic
center.  In order to support themselves against the Galactic potential
with less angular momentum support, the shredded-satellite debris must
then have larger random motions (equivalent to pressure) than do the
typical thick disk stars: this is seen in numerical simulations of
this process (Walker et al.~1996). It is these kinematics which allow
their detection: stars with the highest amplitude of vertical motions
(the satellite debris?) will be preferentially found farther from the
Plane than are most thick disk stars (the heated thin disk?).  If a
population of former satellite stars exists, and the satellite was on
an initial non-circular orbit consistent with cosmological simulations
(van den Bosch et al.~1998), the apparent mean rotational velocity of
stars far from the Plane (the debris) will be less than it is for
stars near the Sun, in the `classical' thick disk.  This situation is
easily distinguished from the possibility that all stars form a
single, coherent, thick disk, in which case the rotational velocities
of the most distant thick disk stars, far from the Plane, will not
differ significantly from those nearby.  This second model is
inconsistent with recent observations.

Determination of the stellar populations in the Galactic thick disk
tests this model, and so constrains the merger history of the Milky
Way, Current indications are that the Galactic thick disk is composed of
only very old stars, ages $\ga 10$~Gyr, equivalent to forming at a
redshift of $\ga 1$ (Wyse 2000).  This implies that the event that
formed it from the thin disk, which now contains stars of all ages,
occurred a long time ago, with little subsequent extraordinary heating
of the thin disk.  If this model is valid, it may be possible to
identify stars captured from the accreted galaxy, and to distinguish
them from those formed in the early thin disk of the Milky Way.  This
would allow tight constraints on what merged, and when it merged, and
on the early star formation in an extended disk.  These are important
tests of hierarchical clustering theories of structure formation.

\begin{figure}
\label{fig8}
\includegraphics[angle=270,width=\columnwidth]{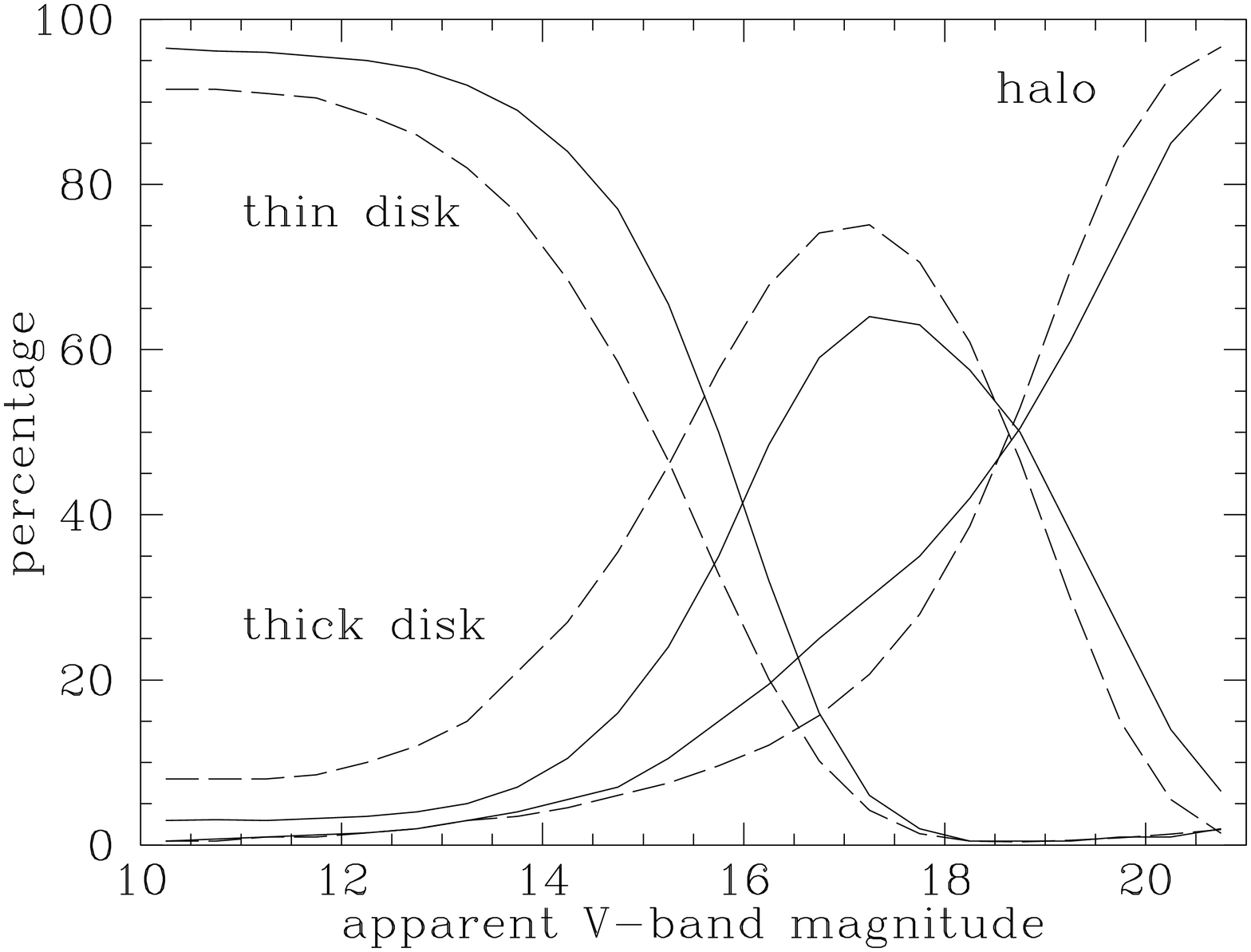}
\includegraphics[width=\columnwidth]{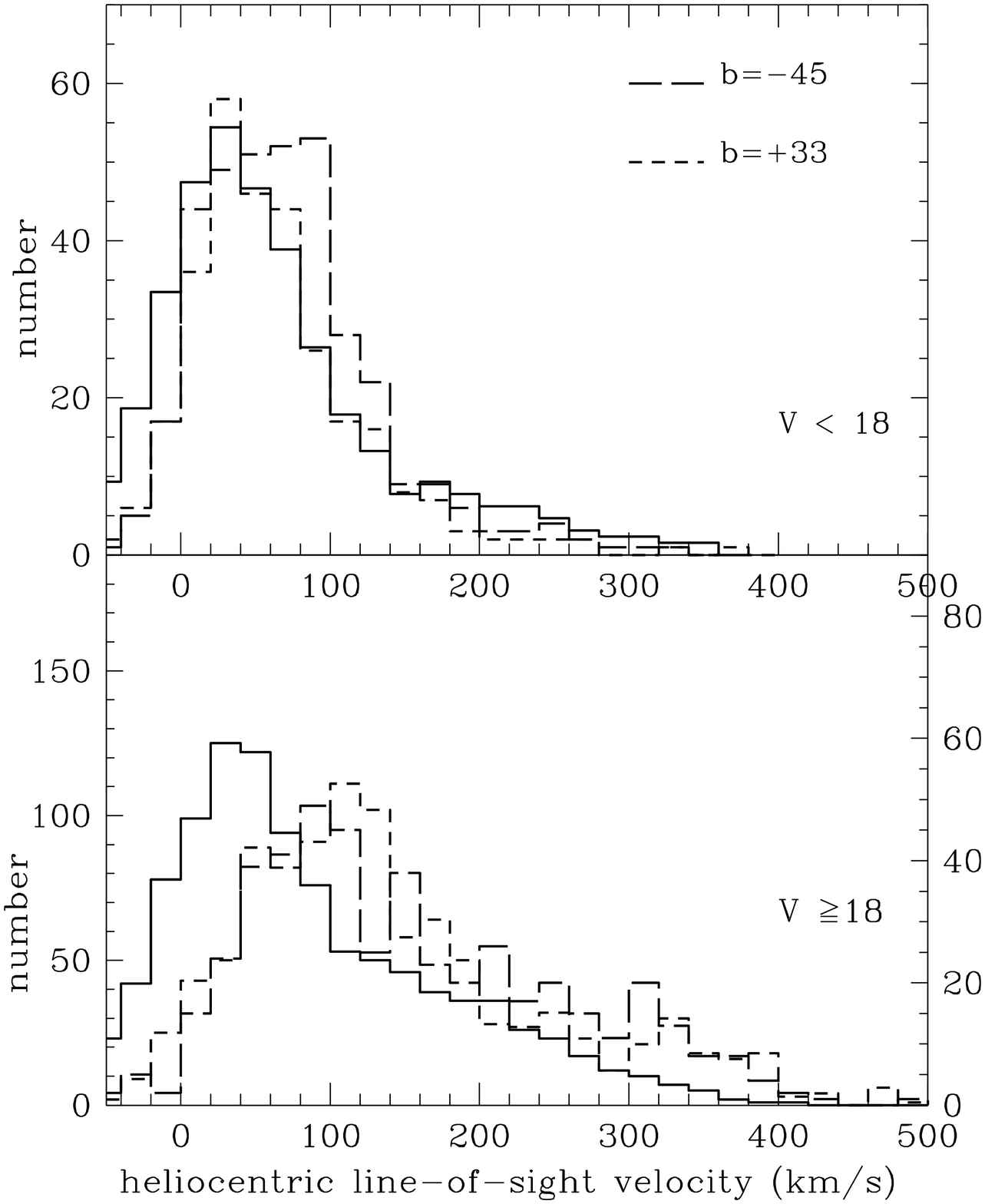}
\caption{TOP: star count predictions as a function of apparent
magnitude, at high Galactic latitudes.
LOWER: Radial velocity histograms for F/G stars in two
lines-of-sight, compared to model predictions.  The kinematics of the
brighter stars, with apparent magnitude less than $18$ in the V-band,
are shown in the top panel and the kinematics of the fainter stars are
shown in the bottom panel.  The solid histograms result from random
sampling Gaussians with `standard' kinematics.
The short-dashed histograms are the data for a field 
in which at these distances the heliocentric line-of-sight
velocity corresponds to $\sim 80$\% of rotation velocity, while the
long-dashed histograms are the data for a second field  in
which it corresponds to $\sim 70$\%.  In the lower panel the y-axis
scale on the right-hand-side refers to the long-dashed histogram. 
Data from Gilmore, Wyse \& Norris 2002.}
\end{figure}

Gilmore, Wyse and Norris (2002; figure~8) have recently
reported first results from a survey with the AAT/2dF designed to
quantify the origins of the thick disk, and to test the merger model
of its formation. They detect a substantial population of stars on
orbits that are intermediate between those of the canonical thick disk
and the canonical stellar halo.

Their data for the brighter stars and the model are in tolerable
agreement showing the well-established canonical thick disk lag of
less than 50km/s. However, there is an obvious disagreement for the
fainter stars, in that the typical star shows a mean lag behind the
Sun of $\sim 100$km/s.  The peak of the observed distribution is
significantly displaced from the model predictions.  This disagreement
is not sensitive to the adopted normalisations or scale heights for
the thick disk and halo, but indicates the need for a substantial
revision in the standard kinematical model of the Milky Way.

Gilmore etal note that they have not detected a small perturbation
superimposed on a smooth well-understood background, but rather
intrinsic complexity in the kinematic distribution function of stars
ascribed in standard models to the thick disk.  They also note that
the predictions of the Gaussian halo fail to reproduce the local peak
in the data at around 300~km/s, which is suggestive of a retrograde
halo stream (velocities above $\sim 180$~km/s in these lines-of-sight
are retrograde), as may be produced by accretion of a small satellite
(e.g.~Helmi et al.~1999).

\section{Conclusions}

A worthwhile galaxy formation algorithm must be able naturally to
reproduce the best observed galaxies, those of the Local Group. The
challenges to the dramatically successful CDM hierarchical model from
the Local Group are demanding, and are currently at the limit of what
the models can predict. Such issues as the number of local dwarf
galaxies and the inner CDM profiles of the dSphs are well known
demanding challenges for CDM models. Other Local group information is
also important: what is the stellar Initial Mass Function, what is the
distribution of chemical elements, what is the age range in the
Galactic Bulge, when did the last significant disk merger happen...
All these challenge our appreciation of galaxy formation and
evolution, and in turn provide the information needed to refine the
models.  The partnership between local observations and ab initio
theory is close, and developing well.

\section{Acknowledgement} It is a pleasure to thanks the organisers
for a wonderful meeting. and my many colleagues whose results I have
summarised here.


\begin{thebibliography}
\bibitem{}d'Antona, {F. 1998}, in `The Stellar {Initial Mass
Function}', ASP Conf.~{Ser. Vol. 142, eds.} G.~Gilmore \& D.~Howell
(San Francisco: ASP{), p. 157} 
\bibitem{} Bahcall, N. {A.}, Ostriker, {J. P.}, Perlmutter, S. \&
Steinhardt, P. {J.}  1999, Sci, 284, 1481
\bibitem{} Baraffe, I., Chabrier, G., Allard, F. \& Hauschildt,
P. {H.} 1997, \aap, 327, 1054

\bibitem{} Blitz L., Binney J., Lo K.Y., Bally J., Ho P.T.P., 1993,
Nature 361, 417 
\bibitem{}Carretta, E., Gratton, R., Cohen, J. G., Beers, T.,
Christlieb, N 2002 AJ 124 481 
\bibitem{}Davidge T.J., 2000, PASP 112, 1177
\bibitem{}Davidge T.J., 2001, AJ 122, 1386
\bibitem{}Davidge T.J., Courteau S., 2002, AJ 123, 1438
\bibitem{}Davies R.L., Frogel J.A., Terndrup D.M., 1991, AJ 102, 1729
\bibitem{}DePoy D.L., Terndrup D.M., Frogel J.A., Atwood B., Blum R.,
1993, AJ 105, 2121 
\bibitem{} Eisenhauer, F., 2001 in `{Starburst Galaxies:} Near and
Far', {eds. L. Tacconi \& D.} Lutz, {(Berlin Heidelberg:
Springer-Verlag), p. 24} 
\bibitem{}Frogel J.A., 1990, in: ESO/CTIO Workshop on Bulges of
Galaxies. ESO (Garching), A92-18101 05-90, p177
\bibitem{} Frogel J.A., 1999a, in: The formation of galactic bulges,
eds.\ C.M.\ Carollo, H.C.\ Ferguson \& R.F.G.\ Wyse. CUP, p38
\bibitem{} Gallagher, {J. S.} \& Wyse, {R. F. G.} 1994, PASP, 106, 1225
\bibitem{}Genzel R., Hollenbach D., Townes C.H., 1994, Rep.Prog.Phys.\ 57, 417
\bibitem{}Gerhard, {O. E.} \& Spergel, {D. N.} 1992, \apjl, {389}, L9
\bibitem{} Gilmore, G. 2001, in `{Starburst Galaxies:} Near and Far',
{eds. L. Tacconi  \& D. Lutz (Berlin Heidelberg: Springer-Verlag), p. 34}
\bibitem{}Gilmore, G. \& Howell, D. {1998, eds.} `The Stellar {Initial
Mass Function}', ASP Conf.~{Ser. Vol. 142 (San Francisco: ASP)}

\bibitem{}  Gilmore, G., Wyse, R. F. G., \&  Norris, J. E.  2002 ApJL 574 39
\bibitem{} Gilmore, G., Wyse, R.F.G. \& Kuijken, K. 1989, ARAA, 27, 555
\bibitem{}Hargreaves, {J. C.}, Gilmore, G., Irwin, {M. J.} \& Carter,
D. 1994, \mnras, 271, 693 
\bibitem{} Helmi, A., White, S.M., de Zeeuw, P. T. \& Zhao, H. 1999, 
Nature, 402, 53
\bibitem{} Hernandez, X., Gilmore, G. \& Valls-Gabaud, D. 2000,
\mnras, {317, 831} 
\bibitem{}  Ibata, R., Gilmore, G. \& Irwin, M. 1994, Nature, 370, 194
\bibitem{} Ibata, R., Wyse, R.F.G., Gilmore, G., Irwin, M. \&
Suntzeff, N. 1997, AJ, 113, 634
\bibitem{} Ibata, R., Irwin, M., Lewis, G. \& Stolte, A. 2001, 
ApJ, 547, L133

\bibitem{} J\o rgensen, I., Franx, M., \& Kj\ae rgaard, P.\ 1996,
\mnras, 280, 167 
\bibitem{} Kleyna, {J. T.}, Wilkinson, M.I., Evans, N.W. \& Gilmore,
G. 2001, ApJL, 563, L115

\bibitem{} Kroupa, P. 2002, Science, 295, 82

\bibitem{}Lamers H.J.G.L.M., Panagia N., Scuderi S., et al., 2002, ApJ
566, 818 
\bibitem{} Madau, P. \& Pozzetti, {L. 2000}, MNRAS, 312, L9
\bibitem{} Majewski, S. 1993, ARAA, 31, 575

\bibitem{} Mateo. M. Annu. Rev. Astron. Astrophys. 1998. 36 435 
\bibitem{} Mateo, M., Olszewski, E. \& Morrison, H. 1998, ApJ, 508, L55
\bibitem{} Meylan, G. 2002, in `Extragalactic Star Clusters', IAU
{Symp.} 207, eds.~E.~Grebel, D.~Geisler \& D.~{Minniti}, p555
\bibitem{} Mighell, {K. J.} \& Burke, {C. J.} 1999, {\aj}, 118, {366}

\bibitem{}Mighell K.J., Rich R.M., 1995, AJ 110, 1649
\bibitem{}Mighell K.J., Corder S., 2002, submitted to AJ
\bibitem{} Muench, A., Lada, E., Lada, C., \& Alves, J., 2002 Ap J 573  366 
\bibitem{}Nakada Y., Onaka T., Yamamura I., 1991, Nature 353, 140
\bibitem{} Olszewski, {E. W.} \& Aaronson, M. 1985, \aj, 90, 2221
\bibitem{} Quinn, P. \& Goodman, J. 1986, ApJ, 309, 472
\bibitem{}Reid, {I. N.}, Kirkpatrick, {J. D.}, Liebert, J., Burrows,
A., Gizis, {J. E.}, Burgasser, A., Dahn, {C. C.}, Monet, D., Cutri,
R., {Beichman, C. A.} \& {Skrutskie},  M. 1999, {\apj, 521}, 613
\bibitem{}
Rejkuba A., Minniti D., Silva D.R., Bedding T.R., 2001, A\&A 379, 781
\bibitem{}
Rich R.M., 1998a, in: The central regions of the Galaxy and galaxies,
ed.\ Y.\ Sofue. Kluwer (Dordrecht), IAUS 184, p11
\bibitem{}
Rich R.M., 2001, in: Astrophysical Ages and Timescales, eds.\ T.\ van Hippel,
N.\ Mansuset \& C.\ Simpson. ASP Conf.Ser.\ 245, p216
\bibitem{}Rich R.M., Mighell K.J., 1995, ApJ 439, 145
\bibitem{} Silk, J. \& Wyse, R.F.G. 1993,  {Physics Reports}, {231}, 295
\bibitem{} Stephens A.W., Frogel J.A., 2002, AJ 124 2023
\bibitem{} Tremaine, S. \& Gunn, {J. E.} 1979, Phys Rev Let, 42, 407
\bibitem{} Van den Bosch, F., Lewis, G., Lake, G. \& Stadel, J. 1999,
ApJ, 515, 50 
\bibitem{} van Loon, J., Gilmore, G., Omont, A., Blommaert, J., Glass, I.,
Messineo, M., Schuller, F., Schultheis, M., Yamamura, I., \&  Zhao,
H-S., 2002 astroph/0210073
\bibitem{} Velazquez, H. \& White, S.D.M. 1999, MNRAS, 304, 254
\bibitem{}von Hippel, T., Gilmore, G., Tanvir, N., Robinson, D. \&
Jones, {D. H. P.} 1996, AJ, 112, 192 
\bibitem{} Walker, I., Mihos, J.C. \& Hernquist, L. 1996, ApJ, 460, 121

\bibitem{}Wyse, R.F.G. 2000, in 35th Liege Colloquium, `The Galactic
Halo, from Globular Clusters to Field Stars', eds A.~Noel et al.,
(University of Liege, Belgium) p305  
\bibitem{}Wyse R.F.G., Gilmore G., Franx M., 1997, ARA\&A 35, 637
\bibitem{} Wyse, R.F.G. \&  Gilmore, G.  1988 AJ 95 1404
\bibitem{} Wyse, R. F. G., Gilmore, G.,  Houdashelt, M., Feltzing, S.,
Hebb, L.,  Gallagher, J. S., III, Smecker-Hane, T., 2002 New Astr. 7 395
\bibitem{}Yanny, B. et al. 2000, ApJ, 540, 825

\end{thebibliography}
\end{document}